\documentclass{svproc}
\usepackage{url}
\usepackage[pdftex]{graphicx}
\usepackage{epstopdf}
\usepackage{framed,multirow}
\usepackage{amsmath}
\usepackage{algorithm}
\usepackage{algorithmic}
\usepackage{amssymb}
\usepackage{subfigure}
\usepackage{booktabs}

\begin{document}
\mainmatter              % start of a contribution
\title{Stochastic Geometric Iterative Method for Loop Subdivision Surface Fitting}
%
%\titlerunning{Hamiltonian Mechanics}  % abbreviated title (for running head)
\author{Chenkai Xu\inst{1} \and Yaqi He\inst{2}\and
Hui Hu\inst{1} \and Hongwei Lin$^*$\inst{,1,2}}
%
%%
%%%%% list of authors for the TOC (use if author list has to be modified)
%Craig Chambers, Kim B. Bruce, and Elisa Bertino}
%%
%\institute{School of Mathematical Sciences, Zhejiang University, Hangzhou, Zhejiang Province, China
%\and State Key Lab. of CAD\&CG, Zhejiang University, Hangzhou, Zhejiang Province, China}
\institute{School of Mathematical Sciences, Zhejiang University, Hangzhou,  China
\and State Key Lab. of CAD\&CG, Zhejiang University, Hangzhou, China
\email{hwlin@zju.edu.cn}
}
%\author{
%    Chenkai Xu \textsuperscript{1[0000-0001-7732-1724]} \and
%    Hongwei Lin \textsuperscript{1,2(\Letter)[0000-0002-9337-9624]}
%    }
%\authorrunning{Chenkai Xu and Hongwei Lin}
%\institute{
%    \textsuperscript{1}    School of Mathematical Sciences, Zhejiang University, Hangzhou, Zhejiang Province, China\\
%    \textsuperscript{2}          State Key Lab. of CAD\&CG, Zhejiang University, Hangzhou, Zhejiang Province, China\\
%    \email{ckxu@zju.edu.cn, (\Letter)hwlin@zju.edu.cn}
%    % \Letter: \email{hwlin@zju.edu.cn}(Hongwei Lin)\\
%  }

\maketitle              % typeset the title of the contribution

\begin{abstract}
%In geometric processing,
%   surface fitting is a popular method for preprocessing the large-scale 3D models.
In this paper,
    we propose a stochastic geometric iterative method
    to approximate the high-resolution 3D models by finite Loop subdivision surfaces.
Given an input mesh as the fitting target,
   the initial control mesh is generated using the mesh simplification algorithm.
Then,
    our method adjusts the control mesh iteratively
    to make its finite Loop subdivision surface approximates the input mesh.
In each geometric iteration,
    we randomly select part of points on the subdivision surface to calculate the difference vectors and distribute the vectors to the control points.
Finally,
    the control points are updated by adding the weighted average of these difference vectors.
We prove the convergence of our method
    and verify it by demonstrating error curves in the experiment.
In addition,
   compared with an existing geometric iterative method,
   our method %a shorter running time and lower error.
    has a faster fitting speed and higher fitting precision.

\end{abstract}
\section{Introduction}\label{sec:1:intro}

In recent years,
    with the improvement of the performance of digital equipment,
    large-scale 3D point clouds and high-resolution triangular meshes have a significant increase in CAD and CAM.
These data often have a large volume and hold noise,
     thus bringing challenges to the related geometric processing algorithms.
A popular method for preprocessing large-scale 3D data is \emph{surface fitting},
    which is designed to approximate or interpolate the input data by optimizing
    the control points of commonly used parametric surfaces,
    such as B-splines
    %NURBS
    and subdivision surfaces.
It makes the fitting surfaces applicable for follow-up studies,
    such as texture mapping,
    iso-geometric simulation and calculation of geometric operators~\cite{XU2021236}.

Geometric iterative methods (GIMs) are a class of iterative methods for surface and curve fitting,
    which have clear geometric meanings and are easy to integrate geometric constraints in the iteration~\cite{lin2018survey}.
In the past 20 years,
    GIMs are successfully employed in geometric design~\cite{lin2018survey},
    data fitting~\cite{Lin2013an},
    reverse engineering~\cite{Lin05curve},
    mesh and NURBS solid generation~\cite{Sasaki2017Adap}.
%The existing GIMs select all data points to update the control points in each iteration,
%   and for the case where the data point parameters cannot be fixed,
%   nearest neighbor search is required.
Although GIM is an effective tool for data fitting,
    there are still slow calculations and even memory overflow when the data size is enormous.
For example,
    in the fast calculation research of the Laplace-Beltrami eigenproblem~\cite{XU2021236},
    for dense meshes,
    surface fitting takes up much time if the iterative algorithm is not optimized.
Therefore,
  %  the GIM should still be improved for efficient large-scale data fitting.
   the GIMs still have the potential to be improved for efficient large-scale data fitting.
 %   it is still necessary to improve the GIM to make it more  for

In this paper,
    we propose the \emph{Stochastic Geometric Iterative Method}(S-GIM) to implement the Loop subdivision surface fitting.
Given a high-resolution mesh,
    our method can quickly generate a finite subdivision surface that approximates the original mesh.
%The idea comes from the stochastic gradient descent in deep learning,
%    where the ``mini-batch'' or ``random-batch'' data is used in training.
Specifically,
     we first generate an initial control mesh through a mesh simplification algorithm.
Then,
    for all data points,
    the nearest neighbor searching is used to determine their foot points on the finite subdivision surface.
In iteration,
   different from the classic geometric approximation algorithms~\cite{Lin15,Morioka08},
    we randomly select ``a batch of foot points'' to compute geometric distances
    (difference vectors)
    and weigh the vectors to update each control point.
Finally,
    we theoretically prove the convergence of our proposed method.
Experimental results show that
    the S-GIM speeds up the iteration and ensures the fitting precision.
Compared with the classic geometric approximation method~\cite{Lin15},
    S-GIM has a shorter running time and lower error under the same number of iteration steps.
In conclusion,
   our method has significant superiority for handling large-scale data fitting problems.

\subsection{Related work}
Optimization fitting and GIMs~\cite{lin2018survey} are the two most classical fitting methods.
The former needs to solve a minimization problem by constructing linear~\cite{HKD2000} or nonlinear~\cite{HDD1994,MK2005} systems.
GIMs,
    including progressive iterative approximation (PIA) methods~\cite{lin2005totally,deng2014progressive,Chen2009}
    and geometric interpolation/approximation methods~\cite{Maekawa07,Morioka08,Lin15},
    define and refine an initial mesh iteratively.
In PIA,
     a one-to-one relationship is established amongst data points and parameter nodes on the fitting surface(curve).
Each parameter node corresponds to a point on the fitting surface(curve),
    which is called \emph{foot point}.
Then,
    the method computes the difference vectors
    between data points and foot points,
    and displaces the control points through the weighted average of the difference vectors iteratively.
%Thus,
%    the iterations in PIA depend on the parametric distance.

On the other hand,
    in some interpolation or fitting tasks,
    foot points with fixed parameters are not determined for data points.
Maekawa et al.~\cite{Maekawa07} proposed the geometric interpolation algorithm for this situation.
On the fitting surface,
    the point closest to the data point is set as the foot point.
Thus,
    the geometric distance, i.e., the distance between a data point and its foot point
    is employed in the iterations of the geometric interpolation algorithm.
Later,
   Nishiyama et al.~\cite{Morioka08} developed a geometric approximation method for the Loop subdivision surface fitting.
Assuming that data points are given on triangular mesh,
   the fitting method first constructs a control mesh by using the QEM mesh simplification~\cite{QEM},
   and then applies the Newton's method to search the foot points on the limit Loop subdivision surface.
Lin et al.~\cite{Lin15} improved the approximation algorithm in the study of building the surface of volume subdivision fitting.
Lin's method uses finite multi-linear cell-averaging (MLCA) surfaces to replace the limit subdivision surfaces,
    then converts the problem of searching the closest point on the parametric surface into searching the closest point on a discrete mesh.
Compared with the Newton's method that needs to derive the gradient,
    using finite subdivision surface can search the closest point faster
    and is also easier to be extended to other parametric surfaces,
    like NURBS.
%Liu et al.~\cite{liu2020neural} proposed another method to determine the foot points in the study of neural subdivision.
%Liu et al.'s method establishes a bijective mapping between the original mesh and the control mesh,
%   and a bijective mapping between the control mesh and the subdivision surface,
%   in the process of simplification and subdivision,
%   respectively.
%Thus,
%    the bijective mapping between the finite subdivision surface and the original mesh
%    can be established.

\section{Methodology}\label{sec:2:method}
In this section,
    the detailed steps of S-GIM are introduced.
At the same time,
    we theoretically prove the convergence of S-GIM.
Through the input of an original high-resolution mesh $M$ (Fig.~\ref{fig:f0}(a)),
    the S-GIM obtains a control mesh $M_0$ (Fig.~\ref{fig:f0}(c)) and its finite subdivision surface $M_k$ (Fig.~\ref{fig:f0}(d)),
    where the $M_k$ approximates the $M$ through iterations.

\subsection{Stochastic geometric iterative method}\label{sec:2.1:S-GIM}
\begin{figure}[!t]
  \centering
    \subfigure[$M$]
         {
        \includegraphics[scale=.3]{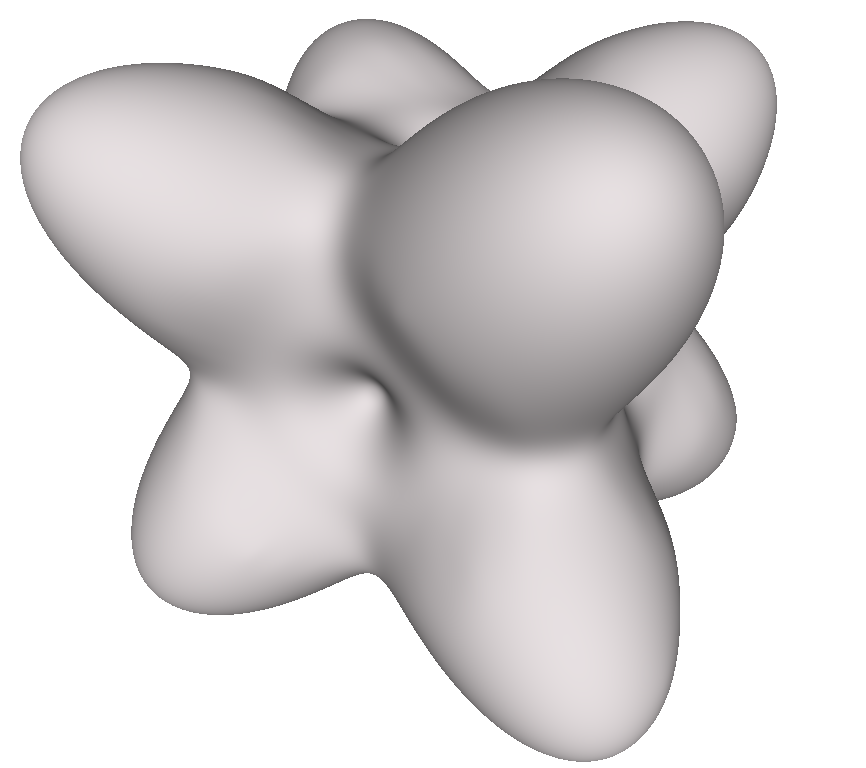}
          }
      \subfigure[$M_0$]
         {
      %   \caption{$M_0$}
         \includegraphics[scale=.3]{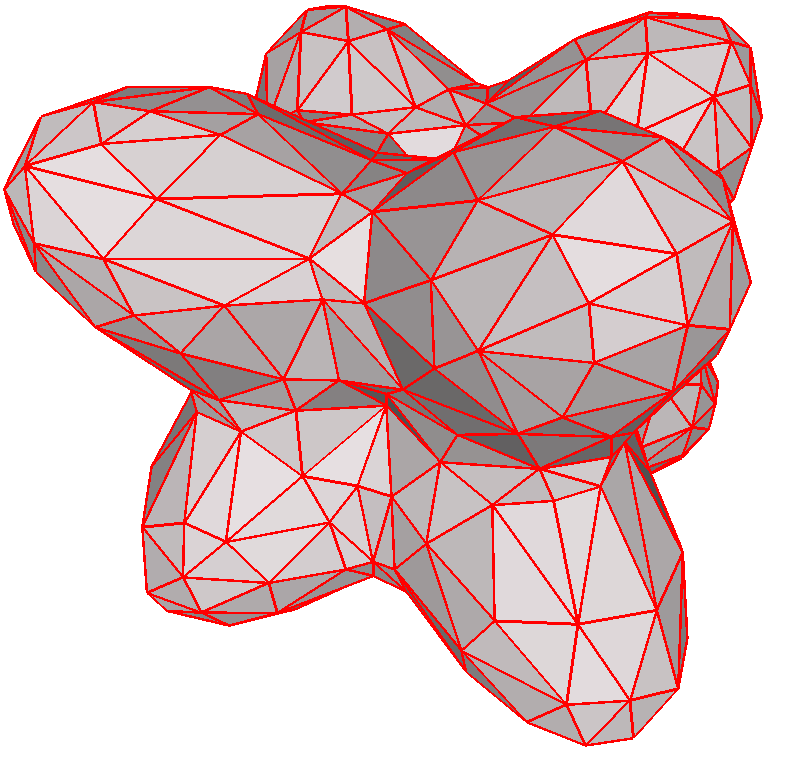}
          }
          \subfigure[$M_0$ after S-GIM]
         {
       %\caption{}
         \includegraphics[scale=.3]{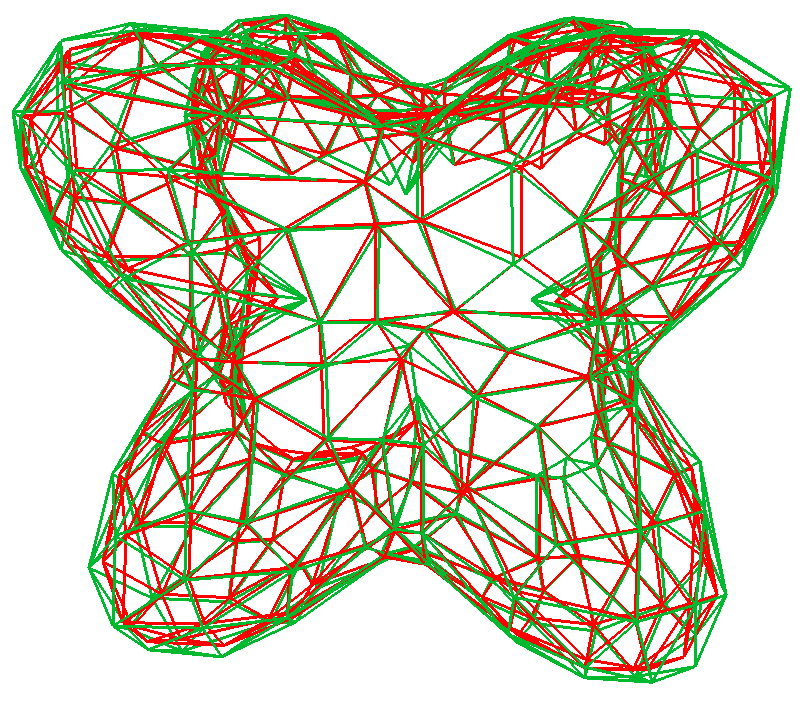}
          }
       \subfigure[$M_k$ after S-GIM]
         {
      %\caption{}
         \includegraphics[scale=.3]{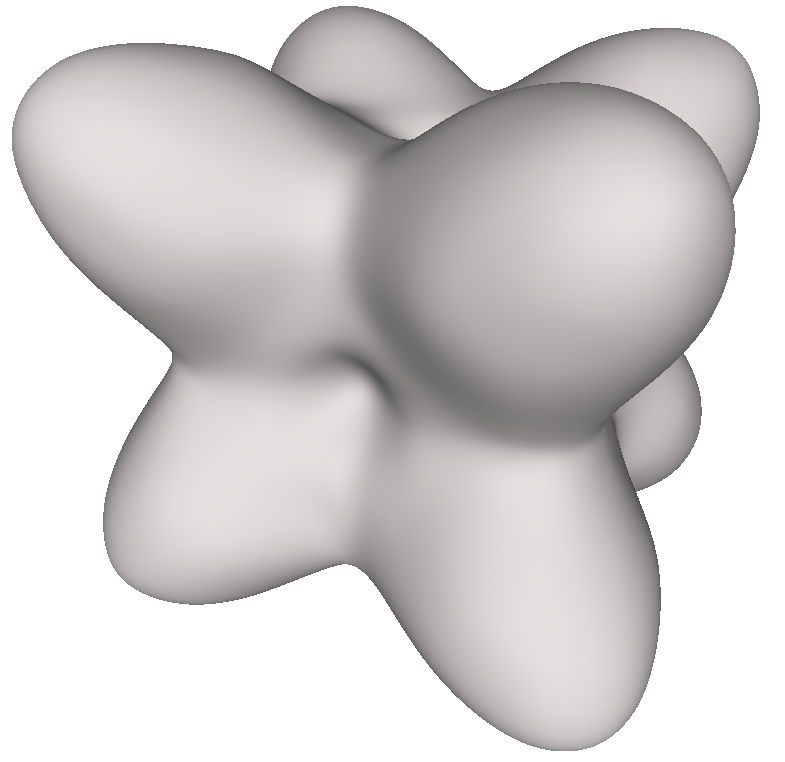}
          }
  \caption{The process of S-GIM for Loop subdivision fitting:
           (a): The input high-resolution mesh $M$ with $160k$ vertices;
           (b): The initial control mesh $M_0$ with 400 vertices;
           (c): The control mesh after iterations;
           (d): The finite Loop subdivision mesh $M_k$ with 26k vertices.
           }
  \label{fig:f0}
\end{figure}

\textbf{Precomputation:}
Before introducing the iteration algorithm,
  the preprocessing work should be carried out in the following steps.

Firstly,
   we use the Loop subdivision surface as the fitting surface because it is easier to obtain the triangular control mesh than regular quadrilateral grids (B-splines).
In this way,
   for complex 3D models,
   we can avoid surface segmentation,
   which is unavoidable in constructing the B-spline (or NURBS) control grids.
In S-GIM,
   the initial control mesh $M_0$ (Fig.~\ref{fig:f0}(b)) is generated using the QEM mesh simplification method.
Referring to the suggestions in~\cite{Morioka08} and~\cite{marinov2005optimization},
   the number of vertices of $M_0$ is set to be $1-5\%$ of $M$,
   which generally leads to good results.

Secondly,
    after fixing the number of subdivision times $k$,
    we can obtain $M_k$ through $k$ times Loop subdivision.
Before the iterations,
    for all points (foot points) on $M_k$,
    their fitting targets (data points on $M$) are determined by nearest neighbor search.
In S-GIM,
    the bounding box space of $M_k$ is divided into $100\times100\times100$.
Then,
    for all data points on $M$,
    we can quickly find their nearest points on $M_k$.
Notably,
    in our experiments,
    we have found that using simple space division can find the nearest points faster than the KD tree method~\cite{muja2009flann}.
Finally,
    for each foot point $p_j \in M_k$,
    we obtain a data point $p_j^* \in M$,
    whose nearest point on $M_k$ is $p_j$.
 If there are several data points whose nearest point is $p_j$,
    we calculate their mean as $p_j^*$.

\textbf{Iteration:}
Let $n_k$ be the number of foot points on $M_k$,
    and $n_0$ be the number of control points on $M_0$.
%    and the subdivision times is set as $k$.
%    the Loop subdivision surface after $k$ times subdivision is $M_k$,
%    which has $m$ vertices.
Each iteration of our method can be expressed as the following steps:
\begin{itemize}
  \item[1.]  Randomly pick $r\%$ foot points, % in $M_k$.
             and denote their index set as $\mathcal{I}=\{a_1, a_2 ,..., a_{n_k*r\%}\}$.
  \item[2.]  For each selected foot point $p_j\in M_k, j\in\mathcal{I}$,
            its fitting target is $p^*_j\in M$.
  \item[3.]  Calculate the difference vector $\delta_j=p^*_j - p_j$,
           and distribute $\delta_j$ to the control points on mesh $M_0$ that generate $p_j$.
  \item[4.] For each control point $v$ on $M_0$,
          average the distributed difference vectors,
          then obtain the difference vector $\Delta$.
  \item[5.] The control point $v$ is updated as $v=\Delta+v$.
  \item[6.] Subdivide $M_0$ $k$ times to produce a new finite subdivision surface $M_k$.
  \item[7.] The above steps are performed iteratively until the fitting precision is reached.
\end{itemize}

The details of the above iteration steps are elucidated in the following.

In the first step,
   not all points on $M_k$ are used to calculate the difference vectors in each iteration.
We call this the ``random batch'' in S-GIM,
  which is borrowed from the conception of ``mini-batch'' in the data training of deep learning.
The number $r$ is called the \emph{sample rate}.
In S-GIM,
    the points of $M_k$ is far more than those of $M_0$,
    that is,
    the variables in fitting (coordinates of all control points on $M_0$)
    are less than the amount of training data.
Therefore,
    it is reasonable to use a part of the data to update $M_0$ in each iteration.
Naturally,
    we need to prove the convergence of our stochastic method in the geometric iterations.
We give a theoretical proof in Section~\ref{sec:proof},
    and show the experimental results for verification in Section~\ref{sec:results}.

In the second step,
   the fitting targets $p^*_j (j=1,...,n_k)$ are updated in the first few iterations.
That is,
   the nearest neighbor search from $M$ to $M_k$ is implemented in these several iterations.
%   to search the nearest point $v_{cl}$,
%   we construct the KD tree of $M_k$,
%   and then search the nearest neighbor of $p_j$ on the KD tree.
%Suppose the number of points of $M_k$ is $m$,
%    then the time complexity of each search is $O(\log m)$.
%Since the number of training points in each step is $N*r\%$.
%The total time complexity of searching in each iteration is $O(m\log m+N\log m*r\%)$,
%    where $O(m\log m)$ is the time to build the KD tree,
%    and $O(N\log m*r\%)$ is the time for searching.
However,
    following the suggestion in ~\cite{Lin15},
    after the first few iterations,
    the fitting targets for $p_j (j=1,...,n_k)$ are fixed on certain points.
And then,
   the time consumption of this step is significantly reduced
   in later iterations.
Other methods for searching the nearest points are given in~\cite{Morioka08} and~\cite{liu2020neural}:
In \cite{Morioka08},
   the Newton's method is employed to calculate the parameter coordinates of the nearest point on the limit surface.
The search time is in the same order of magnitude as in S-GIM.
However,
   compared with our method,
   the Newton's method is more difficult to be implemented and extended to other representation of surfaces (e.g. B-splines).
In \cite{liu2020neural},
    a bijection between $M$ and $M_k$ is established during the preprocessing,
    and then,
    the nearest point of $p_j$ is fixed at the beginning.
Thereby,
     the search time is less than those of our method and the Newton's method.
However,
   the fitting precision is worse than other methods.

In the third step,
   the difference vector $\delta_j$ is defined as
   \begin{equation}\label{eq:delta}
        \delta_j = p^*_j - p_j.
   \end{equation}
It will be distributed to all control points that control the point $p_j$.
Based on the rules of Loop subdivision~\cite{Loop1987},
   $p_j$ is a linear combination of some control points $v_i \in M_0, i=1,...,n_0$,
   i.e.,
   \begin{equation} \label{eq:v sub}
       p_j = \sum^{n_0}_{i=1} c_{ji} v_i, %+ c_2 v^2_0+...+c_n v^n_0,
   \end{equation}
   where the coefficient $c_{ji}$ can be easily obtained from the subdivision process.
Since the subdivision surface is locally controlled by the control points,
   the coefficients are mostly zero.
   %and $\sum_{i=1,..,n} c_i=1$.
Thus,
   the weighted difference vector $c_{ji} \delta_j$ is distributed to the control point $v_i$.
%Since the weights of each control point to calculate $v_{cl}^k$ is recorded in the subdivision matrix~\ref{eq:sub},
%    the time complexity of the second step is $O(m*k)$,
%    where $k$ is the average number of control points that control $v$.

In the fourth step,
  for a control point $v_i$ with a set of weighted difference vectors sequence $\{c_{ji} \delta_j ,j\in\mathcal{I}\}$,
  the difference vector $\Delta$ for $v_i$ is averaged as
    \begin{equation} \label{eq:delta_sub}
    %   \Delta = \frac{\sum_{j\in\mathcal{I}} {c_{ji} \delta_j}}{\sum_{j\in\mathcal{I}} c_{ji}},
     \Delta =\eta \sum_{j\in\mathcal{I}}{c_{ji} \delta_j},
   \end{equation}
   where $\eta$ is the step size,
     which can be adjusted during iteration to ensure convergence.
In this paper,
   the setting of $\eta$ is referenced from the least squares PIA (LS-PIA) method in~\cite{deng2014progressive}.
We set $\eta=\frac{1}{\max\{\sum_{j=1}^{n_k} c_{ji},i=1,\cdots,n_0\}}$,
   which can obtain stable convergence results in the experiment.

In the fifth step,
    $\Delta$ is added to the control point as
       \begin{equation} \label{eq:update}
       v_i=\Delta+v_i.
   \end{equation}
$M_k$ can be updated on the basis of the new control mesh $M_0$ in the sixth step.

Finally,
   to measure the fitting precision,
   we use the root mean square(RMS)
    \begin{equation} \label{eq:rms1}
      e_{r\%}^{(t)}=\sqrt{\frac{\sum_{j\in \mathcal{I}}\|\delta_{j}^{(t)}\|^2}{n_k*r\%}},
   \end{equation}
   or
    \begin{equation} \label{eq:rms2}
      e_{_{100\%}}^{(t)}=\sqrt{\frac{\sum_{j=1}^{n_k}\|\delta_{j}^{(t)}\|^2}{n_k}},
   \end{equation}
   which computes the distance between the subdivision surface $M_k$ and the original mesh $M$ after $t$ times of iterations.
The latter requires more calculation time than the former.
The above steps are performed iteratively until
%     \begin{equation} \label{eq:iter}
%       |\frac{e^{(t+1)}}{e^{(t)}}-1|<\varepsilon_0.
%   \end{equation}
%   In our implementation,
%    we take $\varepsilon_0=10^{-3}$.
    the relative change of the error is less than the preset value,
     i.e.,
     \begin{equation} \label{eq:iter}
       |\frac{e^{(t+1)}}{e^{(t)}}-1|<\varepsilon_0.
     \end{equation}
     or the number of iteration steps reaches a preset value.

We summarize our S-GIM in Algorithm~\ref{alg:S-GIM} and prove the convergence of our method in the next section.

%------------------------------------------------------------------------------
% Algorithm 1
%------------------------------------------------------------------------------
  \begin{algorithm}
  \caption{Stochastic geometric iterative method}
  \label{alg:S-GIM}
  \begin{algorithmic}
  \STATE {\textbf{Input}: original mesh $M$; number of control points $n_0$;
                          number of subdivision times $k$;
                          sample rate $r$}.
  \STATE{$M_0\leftarrow$ QEM$(M,n_0)$};
  \STATE{$M_k\leftarrow$ Loop Subdivision$(M_0,k)$}.
  \WHILE{}%{$ \varepsilon<\varepsilon_0$}
     \STATE{Pick $r\%$ foot points on $M_k$, and build the index set $\mathcal{I}=\{a_1, a_2 ,..., a_{n_k*r\%}\}$};
     \FOR{$j$ in $\mathcal{I}$}
         \STATE{Determine $p_j$'s fitting data $p^*_j$;}
         \STATE{$\delta_j=p^*_j - p_j$};
         \STATE{Distribute $\delta_j$ to $p_j$'s control points};
     \ENDFOR
      \STATE{$e^{(t)}\leftarrow$ Eq.~\ref{eq:rms1} or Eq.~\ref{eq:rms2}}.
      \STATE{If Eq.~\ref{eq:iter} or the steps reaches the preset value, break}.
     \FORALL{$v$ in $M_0$}
      \STATE{$\Delta\leftarrow$ Eq.~\ref{eq:delta_sub}},
      \STATE{$v=\Delta+v$}.
     \ENDFOR
       \STATE{$M_k \leftarrow$ Loop Subdivision$(M_0,k)$};
 \ENDWHILE
 \STATE {\textbf{Output}: $M_k$}.
   \end{algorithmic}
   \end{algorithm}
%------------------------------------------------------------------------------

\subsection{The Proof of the convergence}\label{sec:proof}
In order to prove the convergence of our method,
  we first convert the above iterative steps into a matrix form,
  and then explain the relationship between our method and the stochastic gradient descent,
  that is,
  the special case of the method that randomly selects one foot point in each iteration,
  is same to the standard stochastic gradient descent~\cite{zhang2004}.
Finally,
   following the line in~\cite{zhang2004},
  we prove the convergence of our method that randomly selects a part of the data in each iteration.

\textbf{The matrix form of iteration:  }
In S-GIM,
   the calculations (including difference vectors and all points) on $xyz$-axes are independent.
Thus,
    in the proof,
   we only discuss the case of calculation in one dimension.
In each iteration,
   the control points are the updated object,
   which can be expressed as a matrix as $V=[v_1,v_2,..,v_{n_0}]^T \in \mathbb{R}^{n_0\times1}$.
We denote $V$ at the $t$-th iteration as $V_t$,

At first,
   we discuss the situation of selecting a foot point $p_j$ for fitting.
At this time,
  % the difference $\delta_j$ in Eq.~\ref{eq:delta} is $\delta^{(t)}_j =p^*_j - p_j^{(t)}$.
  according to Eq.~\ref{eq:v sub},
  the difference $\delta_j=p^*_j - p_j$ is distributed to the control points that generate
     \begin{equation*}
        p_{j}=c_{j1}v_1+c_{j2}v_2+...+c_{jn_0}v_{n_0}=C_j^TV_{t-1},
    \end{equation*}
   where $C_j^T=[c_{j1},c_{j2},...,c_{jn_0}]^T \in \mathbb{R}^{1\times n_0}$ is a row vector.
Then,
   by using Eq.~\ref{eq:delta_sub},
   the update of $V_t$ in Eq.~\ref{eq:update} can be expressed as
   \begin{equation}\label{eq:matrix}
       V_t=V_{t-1}+\eta_t(p^*_j-C_j^TV_{t-1})C_j.
   \end{equation}

\textbf{Standard Stochastic Gradient Descent:}
Eq.~\ref{eq:matrix} can be seen as a standard SGD process in~\cite{zhang2004}.
In $t$-th iterations,
    let $(C_j,p^*_j)=(X_t,Y_t)$,
    the coefficients of foot point,
     $C_j$,
      is regarded as the training data $X_t\in \mathbb{R}^{n_0\times1}$,
     the fitting target $p^*_j$ is regarded as the label $Y_t$.
Let the prediction function be $f(X_t,V)=X_t^TV$,
   and the loss function be
     \begin{equation}\label{eq:loss}
       L(f,Y_t)=\frac{1}{2}(Y_t-f)^2.
     \end{equation}
After rewriting Eq.~\ref{eq:matrix} as
    \begin{equation}\label{eq:SSGD}
         \begin{aligned}
              V_t   &=V_{t-1}+\eta_t(Y_t-X_t^TV_{t-1})X_t\\
                    &=V_{t-1}-\eta_t L'_1(X_t^TV_{t-1},Y_t)X_t,\\
         \end{aligned}
   \end{equation}
  where $L'_1(p,q)=\frac{\partial}{\partial p}L(p,q)$,
Eq.~\ref{eq:SSGD} is equal to the unregularized standard SGD~\cite{zhang2004},
    which is the process to solve the optimization problem
       \begin{equation}\label{opt}
             V=\arg\min_W{E_{X,Y}L(W^TX,Y)},
       \end{equation}
    where $X,Y$ is the sample space of data,
    and $E(\cdot)$ is the expectation.
    The convergence of Eq.~\ref{eq:SSGD} has been proven in~\cite{1999Analysis,zhang2004,kivinen2002},
    where a convergence bound for the average loss of a finite sample sequence is obtained.

\textbf{Proof of the convergence of S-GIM:}
In each iteration of S-GIM,
   supposing the size of batch is $m=[r\% \times n_k]$,
   then the training data is $\{X_{t,1}, X_{t,2},...,X_{t,m}\}$.
The iteration of $V_t$ turns into
     \begin{equation}
         \begin{aligned}
             V_t&=V_{t-1}+\eta_t\sum_{i=1}^{m}(Y_{t,i}-V_{t-1}^TX_{t,i})X_{t,i}\\
                &=V_{t-1}-\eta_t\sum_{i=1}^{m}L'_1(V_{t-1}^TX_{t,i},Y_{t,i})X_{t,i}\\
         \end{aligned}
   \end{equation}

Our proof estimates the error boundary between $V_t$ and $V$ according to the method of~\cite{zhang2004},
    which is a more compact analysis method than that of~\cite{1999Analysis,kivinen2002}.
The proof is based on the following decomposition:
   \begin{equation}\label{eq:bound1}
        \begin{aligned}
            \|V_t-V\|^2_{2} &= \|(V_{t-1}-V)-\eta_t\sum_{i=1}^{m}L'_1(V_{t-1}^TX_{t,i},Y_{t,i})X_{t,i}\|^2_{2}\\
                            &= \|V_{t-1}-V\|^2_{2}+\eta_t^2\|\sum_{i=1}^{m}L'_1(V_{t-1}^TX_{t,i},Y_{t,i})X_{t,i}\|^2_{2}\\
                            &-2\eta_t(\sum_{i=1}^{m}L'_1(V_{t-1}^TX_{t,i},Y_{t,i})X_{t,i})^T(V_{t-1}-V)\\
                            &\leq \|(V_{t-1}-V)\|^2_{2}+\eta_t^2\sum_{i=1}^{m}\|L'_1(V_{t-1}^TX_{t,i},Y_{t,i})X_{t,i}\|^2_{2}\\
                            &-2\eta_t\sum_{i=1}^{m}(L(V_{t-1}^TX_{t,i},Y_{t,i})-L(V^TX_{t,i},Y_{t,i})+d_L(V_{t-1}^TX_{t,i},V^TX_{t,i};Y_{t,i})),\\
        \end{aligned}
   \end{equation}
where $\|\sum \vec{a}_i \|^2_{2} \leq \sum\|\vec{a}_i\|^2_{2}$,
   and $d_L(p,q;y)=L(q,y)-L(p,y)-L'_1(p,y)(q-p)$.
$d_L(p,q;y)$ is the Bergman divergence.
For the convex function $L$,
   $d_L(p,q;y)$ is always $\geq0$~\cite{zhang2004},
By using the assumptions in~\cite{zhang2004},
   we have $(L'_1(p,y))^2\leq AL(p,y)+B$,
   the sup of $\|X_t\|^2_{2}\leq M$,
   and $B=0$(the loss function is least square),
   the result of Eq.~\ref{eq:bound1} is
      \begin{equation}\label{eq:bound2}
        \begin{aligned}
             \|V_t-V\|^2_{2}-  \|V_{t-1}-V\|^2_{2}
             &\leq \eta_t^2\sum_{i=1}^{m}\|L'_1(V_{t-1}^TX_{t,i},Y_{t,i})X_{t,i}\|^2_{2}\\
             &-2\eta_t\sum_{i=1}^{m}(L(V_{t-1}^TX_{t,i},Y_{t,i})-L(V^TX_{t,i},Y_{t,i})).\\
             &\leq \eta_t^2AM^2\sum_{i=1}^{m}L(V_{t-1},X_{t,i},Y_{t,i})\\
             &-2\eta_t\sum_{i=1}^{m}(L(V_{t-1}^TX_{t,i},Y_{t,i})-L(V^TX_{t,i},Y_{t,i})).\\
        \end{aligned}
   \end{equation}
Setting $\eta_t=\eta$,
    summing Eq.~\ref{eq:bound2} over $t=1,...,T$
    and rearranging,
    we obtain
     \begin{equation}\label{eq:bound3}
        \begin{aligned}
            &(1-\frac{\eta}{2}AM^2)\frac{1}{mT}\sum_{t=1}^{T}\sum_{i=1}^{m}L(V_{t-1}^TX_{t,i},Y_{t,i})\\
            &\leq \inf_{V} (\frac{1}{mT}\sum_{t=1}^{T}\sum_{i=1}^{m}L(V^TX_{t,i},Y_{t,i}) + \frac{1}{2\eta mT}\|V-V_0\|^2_{2})
        \end{aligned}
        \end{equation}
In S-GIM,
   $\eta$ is small,
   but $\eta mT$ is large,
   then the loss is almost as small as that of the best result of Eq.~\ref{opt}.
This conclusion is the same as Theorem 5.1 in~\cite{zhang2004}(standard SGD),
   which means that the convergence of S-GIM is the same as that in the standard stochastic gradient descent,
   i,e,
   as $T\rightarrow\infty$,
   $E(L(V_T))$ converges to the optimal value if $\eta_t\rightarrow 0$.

\section{Experimental results}\label{sec:results}

In this section,
   the S-GIM for the Loop subdivision fitting is implemented on an i7-9750H 2.60Ghz PC with 16G RAM,
   and compared with a related subdivision surface fitting method~\cite{Lin15}.
The effectiveness of our method is demonstrated in the following figures and tables.

\begin{figure}[!t]
  \centering
     \includegraphics[scale=.5]{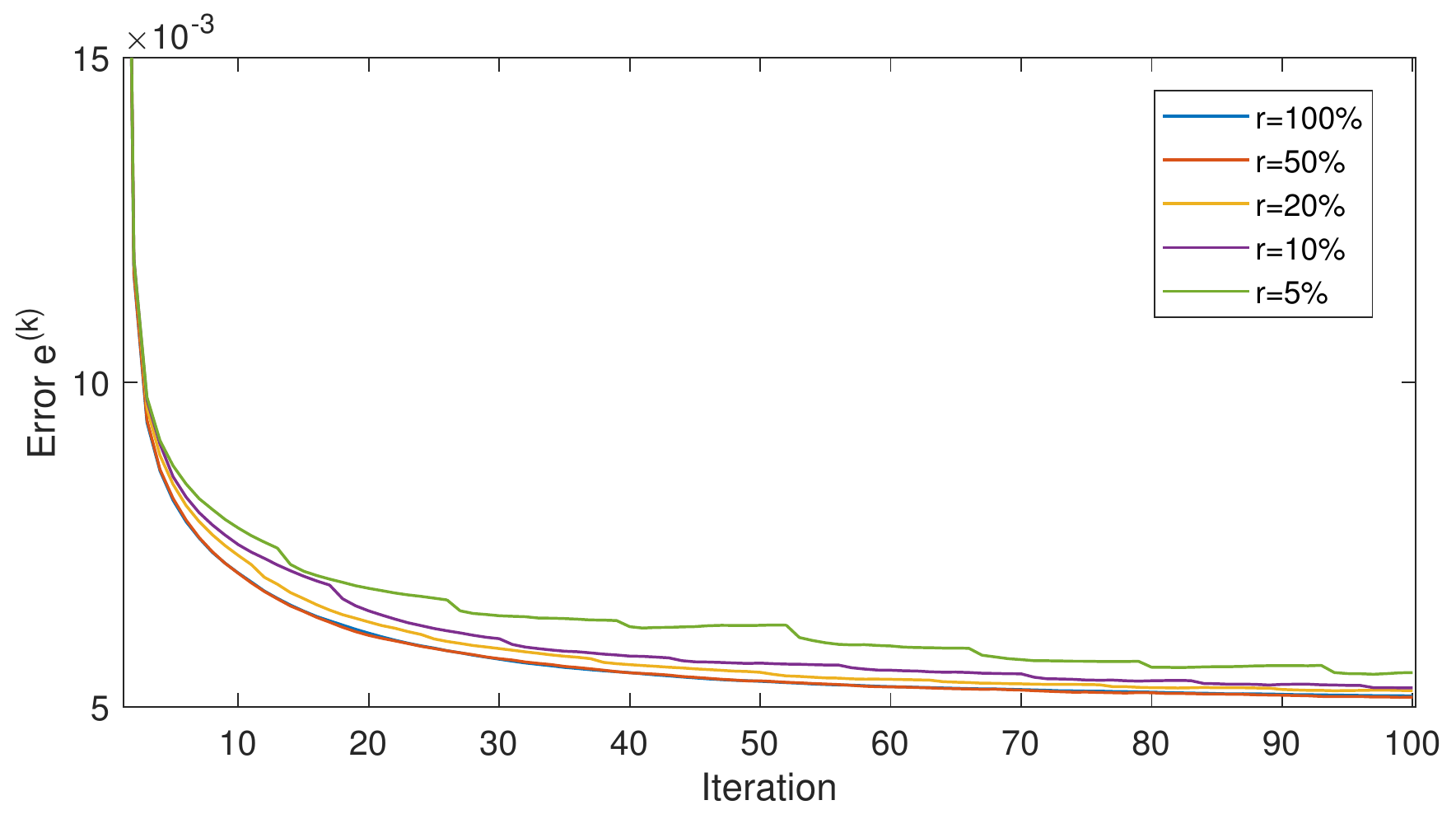}
  \caption{The error curves with different sample rate ($r=5\%, 10\%, 20\%, 50\%~and~100\%$).
            }
  \label{fig:f1}
\end{figure}

In Fig.~\ref{fig:f1},
    the model in Fig.~\ref{fig:f0} is tested.
We set the sample rate $r$ as the experimental variable,
    and draw the error curves under different $r$ values.
The figure shows that,
    when $r=100\%$,
   i.e., all data are used for each iteration,
   the convergence speed of the error curve is fastest.
When $r=50\%$,
    its error curve almost coincides with that of $r=100\%$,
    which means that selecting half of the data for each iteration does not affect the convergence,
    but can reduce the time consumption of each iteration.
Then,
    we further reduce the sample rate,
   when $r=20\%$ and $r=10\%$,
   the convergence speeds are slower than those of $r=50\%$ and $r=100\%$.
However,
    the curves still converge to a value close to $r=50\%$ and $r=100\%$.
Finally,
   when $r=5\%$,
   the error curve converges to an error which is larger than others at the 100-th steps,
   and the curve fluctuates (error increases at some steps),
   which will cause mistakes in judging whether to stop the iteration based on Eq.\ref{eq:iter}.
   So it is inappropriate to choose a lower sample rate for S-GIM.
%The above results verify the proof of the convergence in Section~\ref{sec:proof}:
%As the sample size increases,
%   the difference $\widetilde{\Delta}^{(\gamma)}_i$ is closer to the expected value $\Delta^{(\gamma)}_i$.
%When the selected sample size is small ($r=5\%$),
%   the difference between $\widetilde{\Delta}^{(\gamma)}_i$ and $\Delta^{(\gamma)}_i$ will be large,
%   and then,
%   the error curve will converge slowly or even be non-convergent.
Fig.~\ref{fig:f1} discusses the convergence of our method from the aspect of experiment,
    and provides a reference for selecting the parameter $r$.
In our experiments,
    $r$ is set to $10 - 20\%$,
    which can ensure that the error is close to that of $r=100\%$
    and make the iterations as fast as possible.

 \begin{figure}[!t]
  \centering
     \includegraphics[scale=.45]{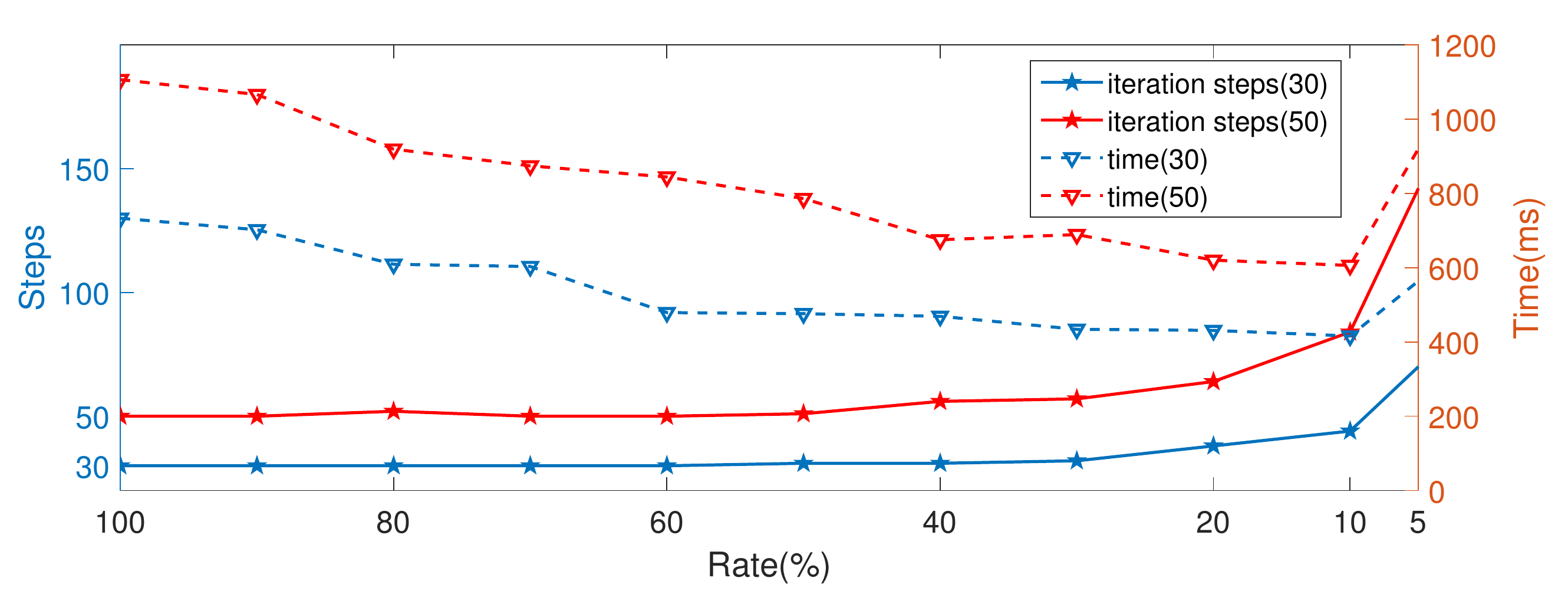}
  \caption{The steps and time consumption to reach the preset errors (the errors in 30-th and 50-th step when $r=100\%$) under different sample rates.
            }
  \label{fig:f2}
\end{figure}

In Fig.~\ref{fig:f2},
 we demonstrate the number of iteration steps and time consumption required to reach the preset error under different sample rates.
Let $e^{(30)}_{100\%}$ and $e^{(50)}_{100\%}$ be the errors in the 30-th and 50-th step respectively,
   and we preset them as the expected errors.
Then our algorithm is tested under varying sample rates.
The iteration is stopped when the error reaches the preset $e^{(30)}_{100\%}$ or $e^{(50)}_{100\%}$.
The ``rate-time" and ``rate-step" curves are displayed in Fig.~\ref{fig:f2}.
The blue solid and dotted line set $e^{(30)}_{100\%}$ as the expected error.
The red solid and dotted line set $e^{(50)}_{100\%}$ as the expected error.
It is shown in Fig.~\ref{fig:f2} that,
   as the sample rate decreases,
   the number of iteration steps required to reach the preset error increases.
However,
    when $40\%<r<100\%$,
    the change in the number of steps is not evident,
    thereby corresponding to the result in Fig.~\ref{fig:f1},
    that is,
    the error curves of $r=50\%$ and $r=100\%$ are close.
In addition,
   although the number of steps is increasing,
   the total iteration time is decreasing.
Except when $r=5\%$,
   the number of iteration steps increases largely,
   and the total calculation time increases.
In conclusion,
   choosing a proper sample rate can reduce the iteration time and ensure the error,
   thus setting $r=10-20\%$ is the best choice.

\begin{table*}[!t]
\centering
\begin{tabular}{|p{2.8cm}|p{1.2cm}|p{1.2cm}|p{0.8cm}|p{1.7cm}|p{1.1cm}|p{1.0cm}|p{1.3cm}|p{1.2cm}|}
\hline
\hline
  \multirow{2}{*}{Model} &\multirow{2}{*}{\shortstack{Original\\mesh}} &\multirow{2}{*}{\shortstack{Control\\mesh}}&\multirow{2}{*}{\shortstack{Sub.\\times}} & \multirow{2}{*}{Error($10^{-3}$)}
  &\multicolumn{4}{|c|}{Time(s)}       \\
  \cline{6-9}
  & & & & &QEM &Loop &Iteration &Total\\
  \hline
  \hline
  \multicolumn{9}{|c|}{Ours} \\
\hline
   Cube8(Fig.~\ref{fig:f0},~\ref{fig:f1}~\&~\ref{fig:f2})    &   163k   & 0.4k &3  & \textbf{6.423} &81.02\% &2.90\%& 16.07\%& \textbf{5.033} \\
\hline
    Bunny(Fig.~\ref{fig:f3})  & 37k   & 0.3k   &3 & \textbf{3.889} &75.82\%& 5.51\% & 18.67\%& \textbf{1.034}  \\
\hline
   Armadillo   & 221k    & 2k   &2  & \textbf{161.328} &48.29\% &0.85\%&50.87\% & \textbf{13.799}\\
   \hline
   Buddha  &543k    & 7k   &2    &\textbf{0.957}  &69.45\%& 1.69\%&28.85\% & \textbf{13.568}     \\
\hline
\hline
  \multicolumn{9}{|c|}{Results in~\cite{Lin15}} \\
\hline
     Cube8(Fig.~\ref{fig:f0},~\ref{fig:f1}~\&~\ref{fig:f2})   &   163k   & 0.4k &5  &  7.671 &20.26\%& 8.67\%& 71.06\% & 20.124     \\
\hline
    Bunny(Fig.~\ref{fig:f3})  & 37k   & 0.3k   &4    & 5.277  & 36.28\%& 6.34\%&57.38\% &2.161 \\
\hline
   Armadillo   & 221k    & 2k   &4  & 166.851 &21.23\%& 6.49\%& 72.28\% & 31.384   \\
   \hline
   Buddha  &543k    & 7k   &3    & 1.121 & 26.12\%& 3.12\%&70.75\%  & 36.073    \\
\hline
\hline
\end{tabular}
\caption{Comparisons with other method.}
\label{tbl1}
\end{table*}

\begin{figure}[!t]
  \centering
        \subfigure%[Cube8.obj: $M$]
         {
         \includegraphics[scale=.16]{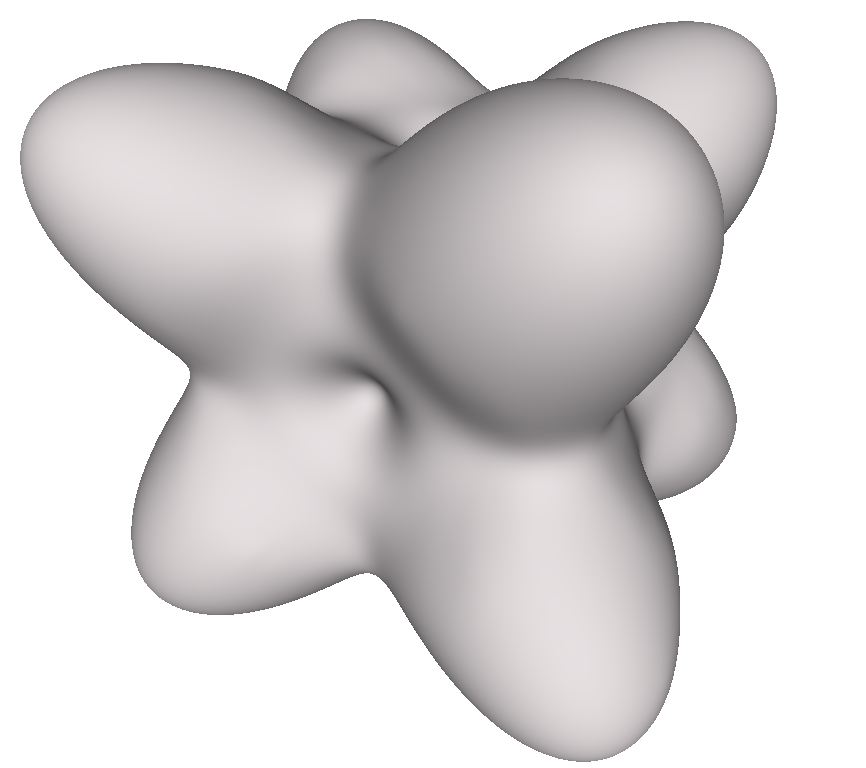}
         \includegraphics[scale=.21]{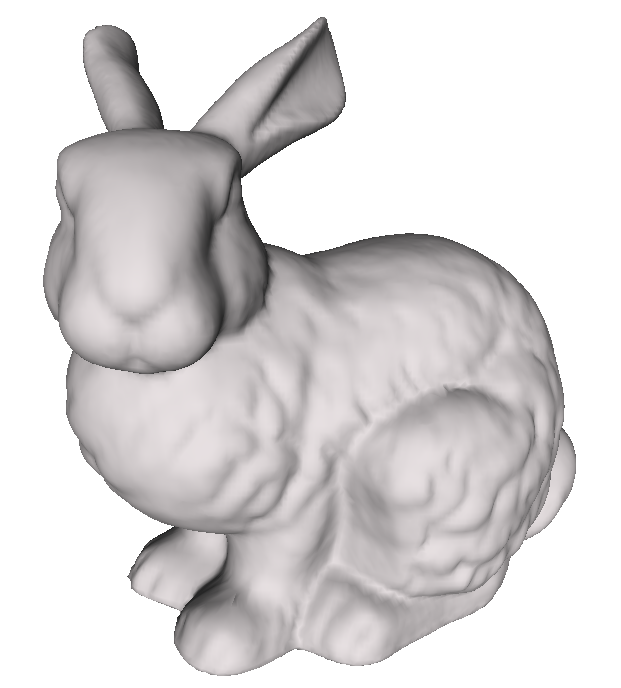}
         \includegraphics[scale=.21]{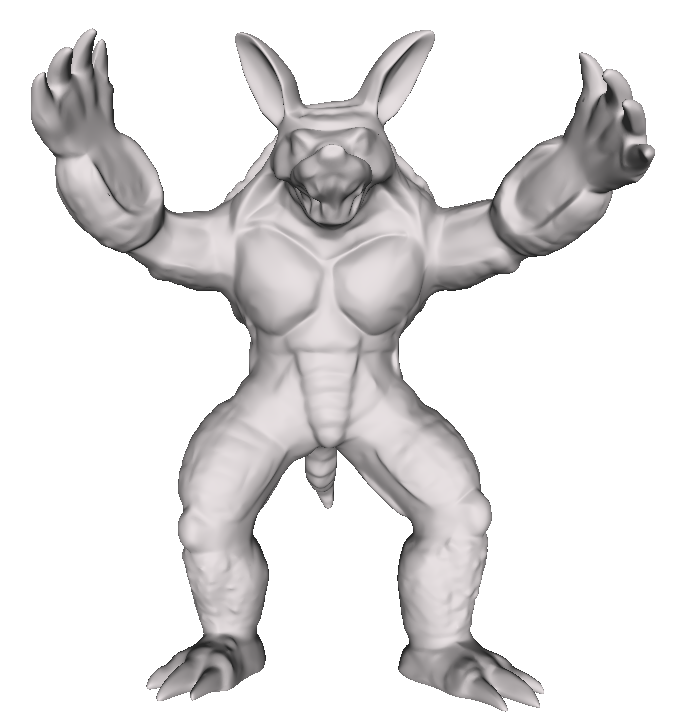}
         \includegraphics[scale=.21]{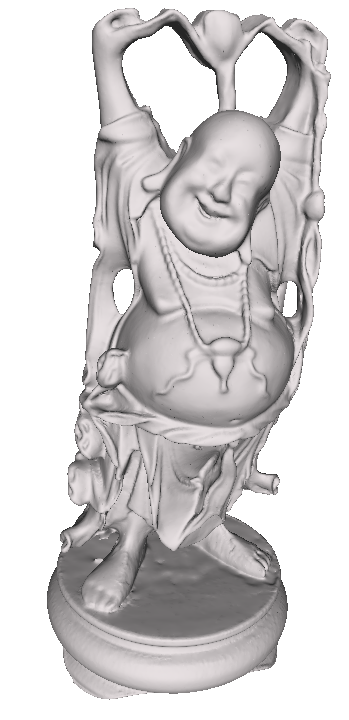}
          }
           \subfigure
         {
          \includegraphics[scale=.16]{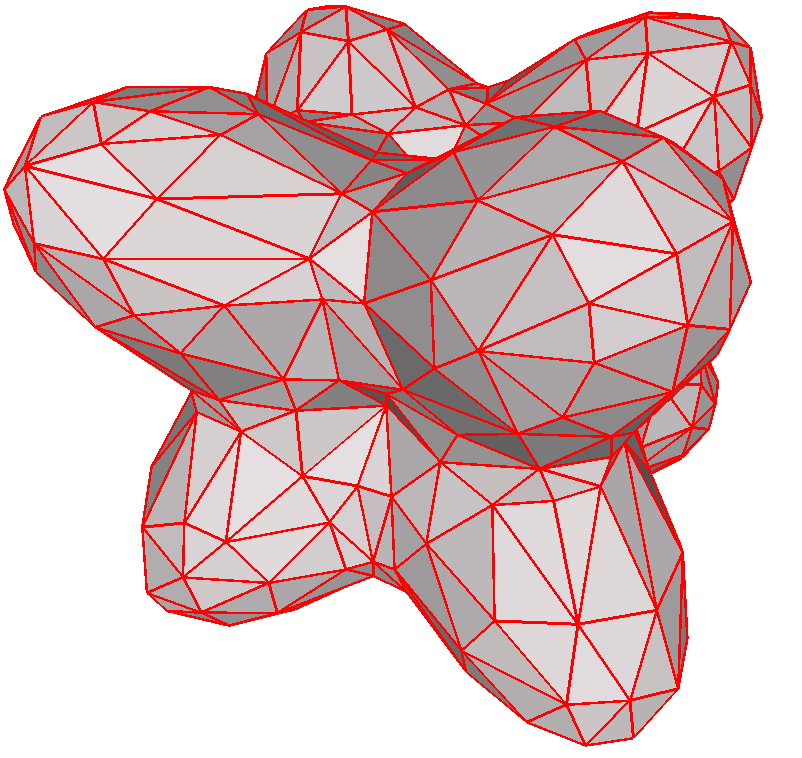}~~~
          \includegraphics[scale=.24]{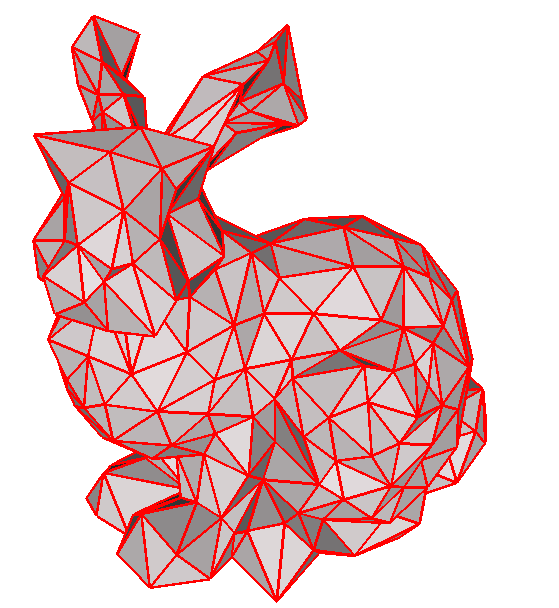}~~
          \includegraphics[scale=.21]{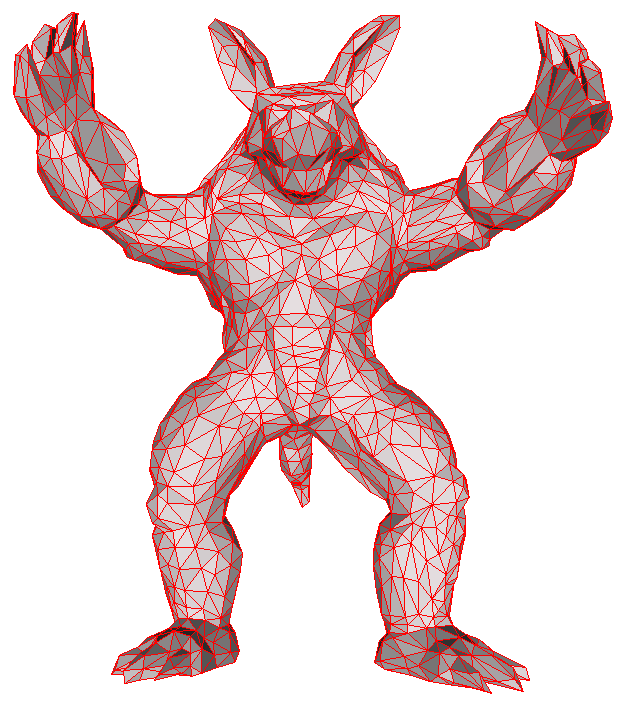}~~
         \includegraphics[scale=.21]{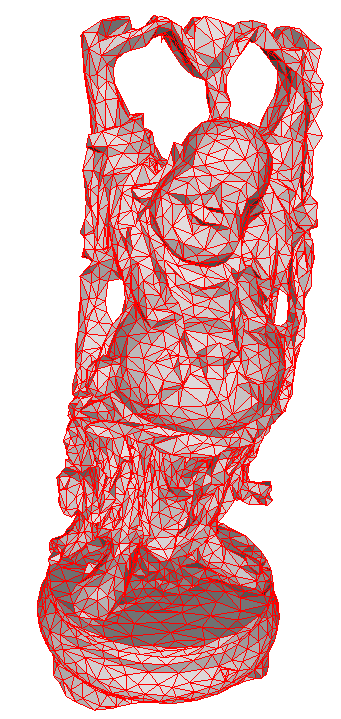}
        }
        \subfigure[Cube8.obj]
         {
          ~~~\includegraphics[scale=.16]{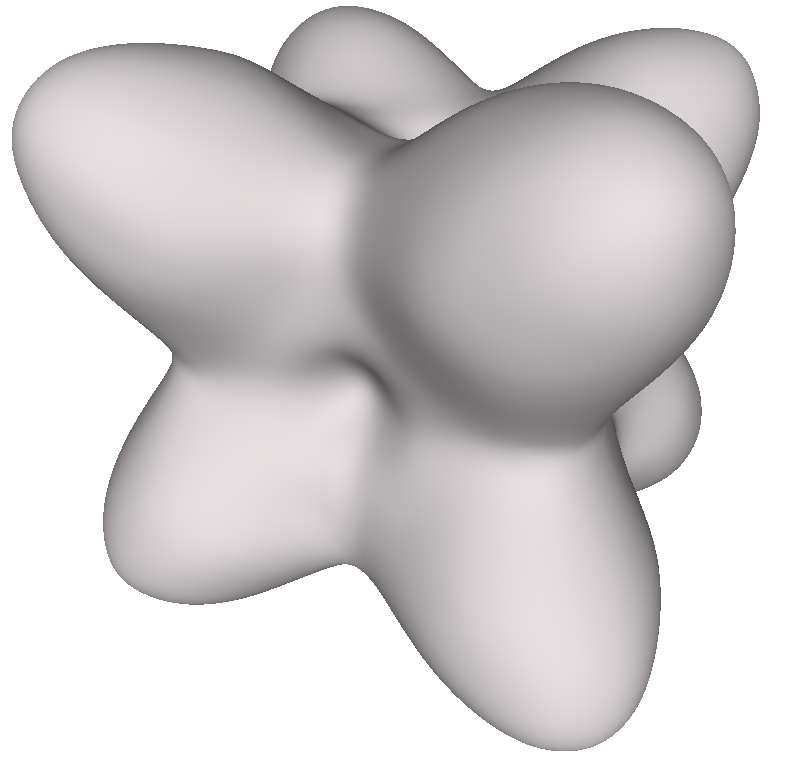}~~
           }
        \subfigure[Bunny.obj]
         {
          \includegraphics[scale=.21]{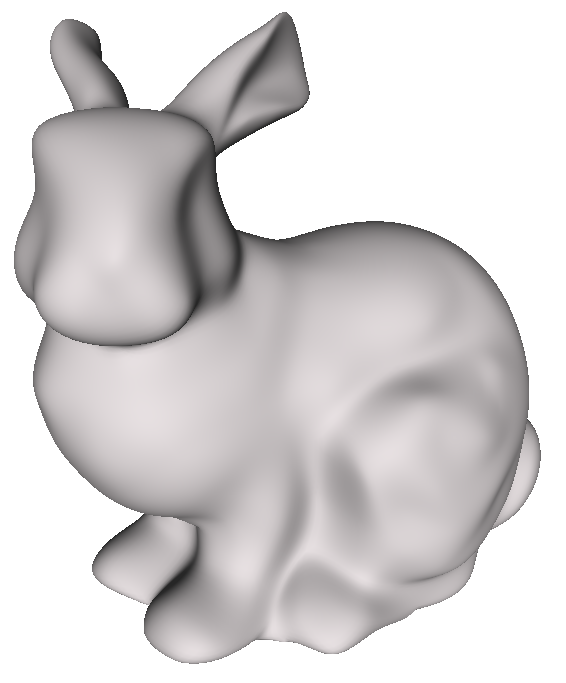}~~
         }
        \subfigure[Armadillo.obj]
         {
          \includegraphics[scale=.21]{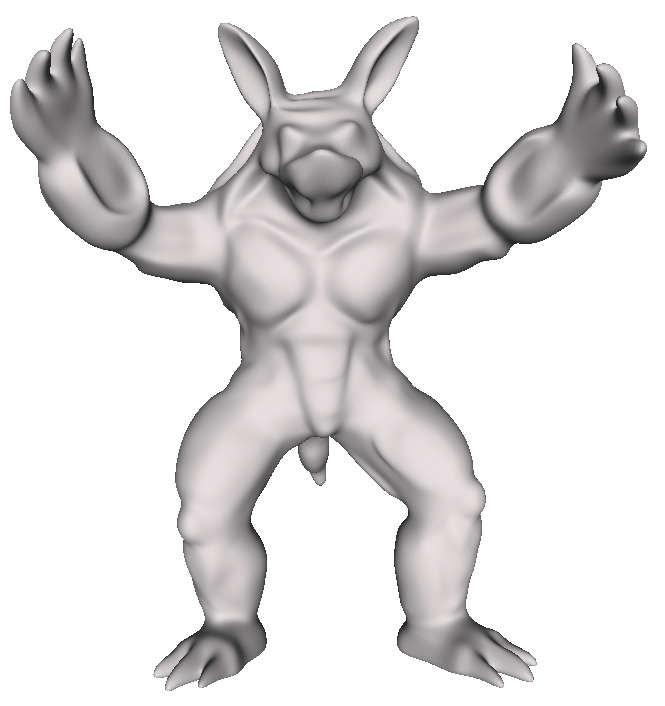}~~
         }
        \subfigure[Buddha.obj]
         {
         \includegraphics[scale=.21]{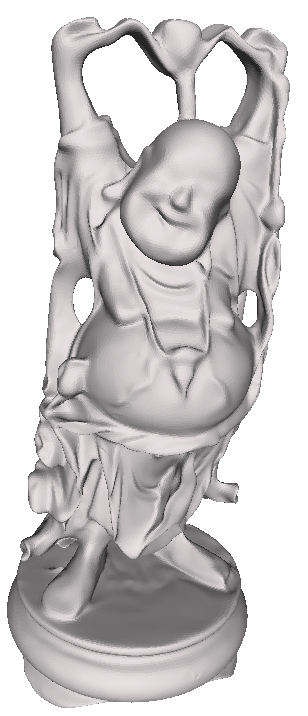}
         }
  \caption{The fitting results of all models tested in Table~\ref{tbl1}.
           The first row is the input large-scale models $M$,
           the second row is the control mesh $M_0$ after iteration,
           and the third row is the subdivision surface $M_k$ that generated by $M_0.$}
  \label{fig:fall}
\end{figure}

Finally,
    we list the test parameters and results in Table~\ref{tbl1},
    and compare them with the \emph{progressive volume subdivision fitting} in~\cite{Lin15}.
Meanwhile,
    the four models tested in the Table~\ref{tbl1} with their fitting results are shown in Fig.~\ref{fig:fall}
In all experiments,
    the number of iteration steps is fixed to 50,
    then,
     the errors and time consumption of all models and different methods at 50 steps can be compared.
Results show that,
     for each tested model,
     the error of our method is smaller than that of the method in~\cite{Lin15},
     and that the running time of our method is at least twice faster than that of~\cite{Lin15}.

\section{Conclusion}\label{sec:4:concludes}
In this paper,
   we propose the S-GIM method to fit large-scale meshes by Loop subdivision surfaces.
Our method uses part of the data in each iteration to calculate the difference vectors and update the control points.
At the same time,
   we prove that S-GIM is convergent.
Experimental results show that when an appropriate sample rate($r=10\% - 20\%$) is selected,
   S-GIM can converge to the expected error and accelerate the convergence speed.

\begin{figure}[!t]
  \centering
    \subfigure[$M$]
         {
        \includegraphics[scale=.3]{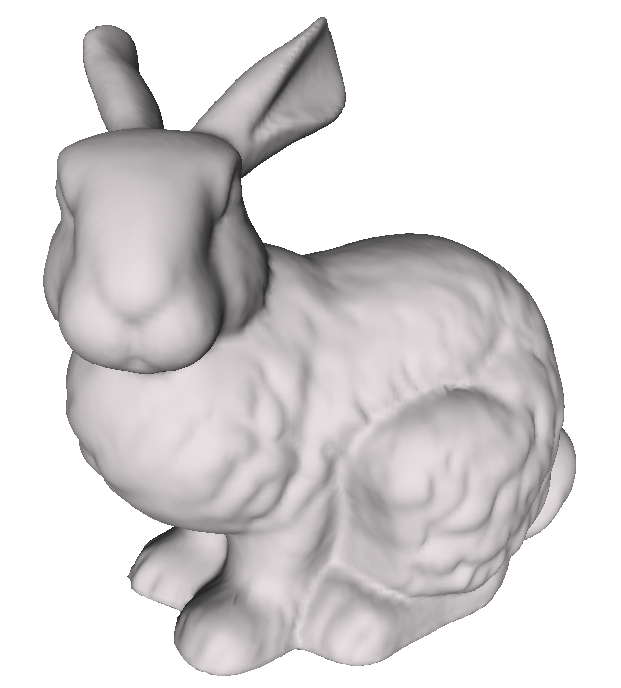}
          }
      \subfigure[$M_k$ with $300$ vertices]
         {
      %   \caption{$M_0$}
         \includegraphics[scale=.3]{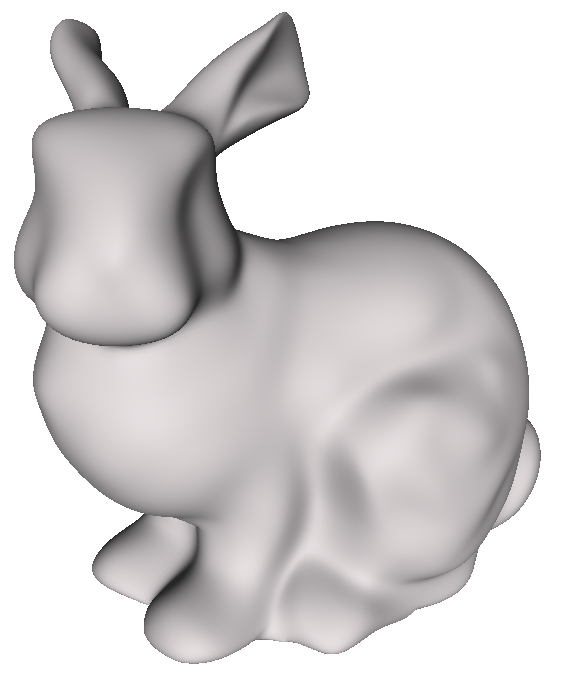}
          }
          \subfigure[$M_k$ with $7k$ vertices]
         {
       %\caption{}
         \includegraphics[scale=.3]{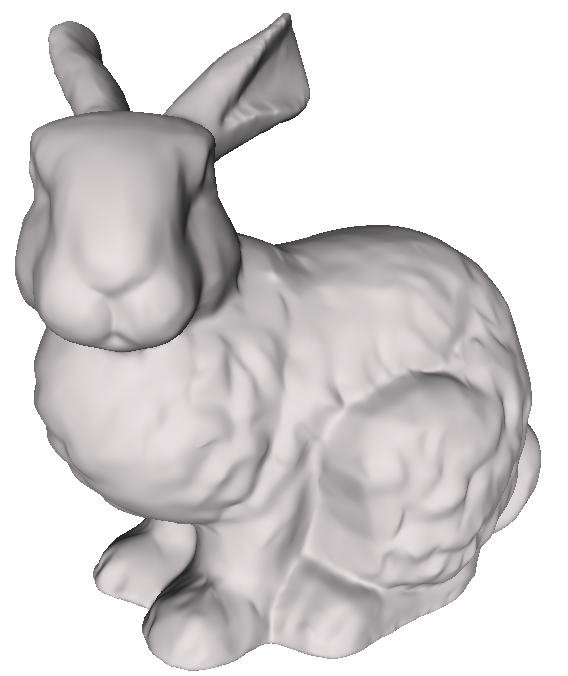}
          }
  \caption{Increasing the control points can improve the quality of the fitted mesh.
           }
  \label{fig:f3}
\end{figure}

For future works,
   in order to further reduce the fitting error,
   we consider adding the incremental data fitting method~\cite{deng2014progressive} into S-GIM.
   %,which is also called the topological adjustments in~\cite{Morioka08}.
   %the topological adjustments proposed in~\cite{Morioka08} into S-GIM.
In Fig.~\ref{fig:f3},
   the results show that increasing the control points can improve the quality of the fitted mesh.
The incremental method inserts new control points during iterations.
The results in \cite{Morioka08} show that inserting control points during iterations
    improves the fitting precision
    while increases the time consumption.
%than that of adding control points at the beginning.
%The method in \cite{Morioka08} inserts new control points in the patch with a large error during iterations.
%Inserting control points during iterations have lower fitting error than adding control points at the beginning.
%However,
%   inserting control points significantly increases the time consumption.
In addition,
   extending the S-GIM to other forms of parametric surface fitting problems,
   such as the geometric approximation of B-spline surfaces or NURBS surfaces,
   is also worthy of study.
As long the amount of fitting data points is far higher than the number of control points that need to be adjusted,
 the S-GIM is feasible and easy to be promoted.

%\section*{Acknowledgments}
%Acknowledgments should be inserted at the end of the paper, before the
%references, not as a footnote to the title. Use the unnumbered
%Acknowledgements Head style for the Acknowledgments heading.
%
%
%
\bibliographystyle{splncs04}
\bibliography{ref}

\end{document}